\documentclass[11pt,twoside]{article}
\usepackage{Cargesepasp}
\usepackage{graphicx}
\def\edcomment#1{\iffalse\marginpar{\raggedright\sl#1\/}\else\relax\fi}
\reversemarginpar

\begin{document}

\title{The Low-Mass Stars in Starburst Clusters}

\author{Bernhard Brandl, the NGC\,3603\footnotemark[1], and the
  30\,Doradus-Team\footnotemark[2]}
\affil{Cornell University, 222
  Space Sciences Building, Ithaca, NY 14853, USA}

\begin{abstract}
  If star formation depends strongly on the global properties of the
  environment in which they form, the slope of the IMF as well as the
  upper and lower mass cut-offs would be expected to vary
  significantly between violent starbursts and more quiescent star
  forming regions.  To test this hypothesis we observed the stellar
  content of the two closest massive star forming regions, 30~Doradus
  and NGC~3603 with HST/NICMOS and VLT/ISAAC.  Our observations are
  the most sensitive observations made to date of dense starburst
  cores, allowing us to investigate its low-mass stellar population
  with unprecedented quality.  The NIR luminosity function of
  30~Doradus shows {\em no} evidence for a truncation down to at least
  $1 M_\odot$, its stellar core R136 is populated in low-mass stars to
  at least $2 M_\odot$ and NGC~3603, which is very similar to R136,
  down to $0.1 M_\odot$.  Our observations clearly show that {\em low-mass
  stars do form in massive starbursts}.
\end{abstract}
\footnotetext[1]{W. Brandner, F. Eisenhauer, A.F.J. Moffat, F. Palla
  \& H.  Zinnecker} 
\footnotetext[2]{H. Zinnecker, W. Brandner, A.
  Moneti, D. Hunter, M. McCaughrean, G. Meylan, R. Larson, M. Rosa, N.
  Walborn \& G. Weigelt}


\section{Introduction}
The stellar initial mass function (IMF) [Salpeter (1955), Miller \&
Scalo (1979), Scalo (1986), Kroupa, Tout \& Gilmore (1990), ...] is
the distribution of stellar masses at $t=0$.  It is mainly
characterized by three parameters: the upper mass cutoff $m_u$, the
lower mass cutoff $m_l$, and the slope $\Gamma = \frac{d\log \xi (\log
  m)}{d \log m}$.  The values of $\Gamma, m_u,\mbox{\ and\ }m_l$
depend on the physical processes that lead to the formation of stars
and hence may not be ``universal'' but depend rather significantly on
the environment, i.e., the global parameters of the region.

An increasing number of theoretical and semi-empirical approaches try
to explain the stellar mass spectrum, e.g., Silk (1977), Padoan et al.
(1997), Murray \& Lin (1996), Elmegreen (1998), Adams \& Fatuzzo
(1996), Bonnell et al. (1997), Myers (2000), and recent models
discussed at this conference.  Despite numerous and diverse approaches
none of the concepts seems to be capable of predicting the stellar
mass spectrum under the extreme conditions in starbursts.  Hence,
observational input is needed!

Evidence for a different IMF in star-forming regions has been found in
several observational studies: the element abundances in spiral
arms/interarm regions ($m_l \ge 2M_\odot$ [G\"usten \& Mezger 1982]),
the metal abundances in galaxy clusters ($\rightarrow$ top-heavy IMF),
cooling flows in galaxy clusters ($\rightarrow$ bottom-heavy IMF), and
studies of isolated LMC/SMC associations ($\Gamma\approx 3.5$ [Massey
et al.  1995]).  Maybe the strongest support for a substantially
higher low-mass cut-off in starbursts came from a comparison of
radiation {\em and} gravitation for M82, suggesting $m_l \sim
3-5M_\odot$ (Rieke et al. 1993).  However, starburst modelling is a
tricky business and depends on various assumptions, and a similar
analysis of M82 by Satyapal et al. (1995) did not confirm the high
value of $m_l$.

\begin{figure}\centering
\includegraphics[scale=0.4]{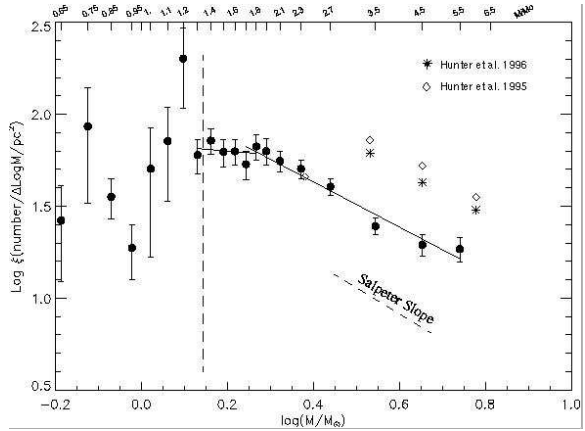}
\caption{\label{sirianni} A ``definite flattening
  below $\sim 2 M_\odot$'' of the IMF in 30~Doradus was found by
  Sirianni et al. (2000), based on very deep V and I broadband WFPC2
  images from the HST archive.}
\end{figure}

Recently, Sirianni et al. (2000) have investigated the low-mass end of
the IMF around R136 based on very deep WFPC2 images in V and I
band from the HST archive and extended the mass range below Hunter et
al.'s (1995) limit of $2.8 M_\odot$.  The authors find that ``after
correcting for incompleteness, the IMF shows a definite flattening
below $\sim 2 M_\odot$'' (Fig.~\ref{sirianni}).  However, their
conclusion is based only on regions $r \ge 1.5$\,pc {\em around}
R136 and it remains unclear whether the conditions there are still
representative of dense starburst cores. On the other hand, most stars
less massive than $\sim 2-3 M_\odot$ show varying amounts of
extinction due to circumstellar disks or envelopes, which --- if the
observations are performed at optical wavelengths --- require a
magnitude-limit correction (Selman et al. 1999) in addition to
correction for incompleteness.

\begin{figure}\centering
\includegraphics[scale=0.55]{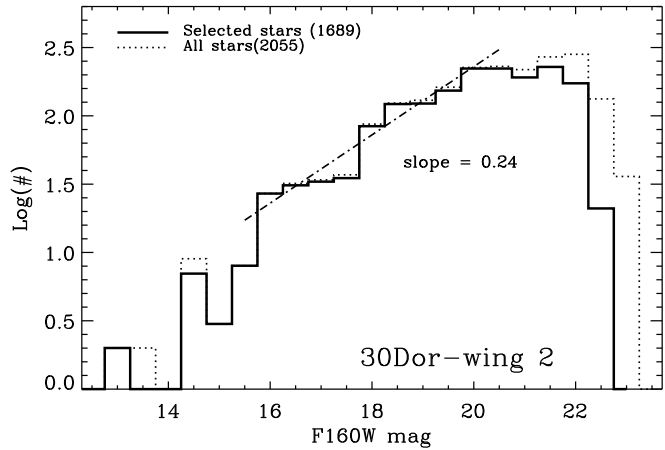}
\caption{\label{nic2lumi} Infrared luminosity function (F160W, NIC2) 
  for ``wing~2'' in 30~Dor.  `Selected stars' refers to stars with
  a magnitude error $\le 0.2$~mag were considered (solid line)
  [Zinnecker et al. 1998].}
\end{figure}

Most observational methods used to study starburst regions are based
on nebular emission lines, far-IR fluxes, UV stellar winds, and red
supergiant spectral features --- diagnostics which are most sensitive
to OB stars with little information on the low-mass star content.  The
best way to get an accurate census of the entire stellar population is
through the detection of individual stars by means of high angular
resolution techniques at near-IR wavelength.  The near-IR is the ideal
wavelength range for these studies since extinction is significantly
reduced ($A_K \sim 0.11 A_V$, Rieke \& Lebofsky [1985]) and the
difference in luminosity between young stars spanning three orders of
magnitude in mass is reduced by a large factor compared to optical
wavelengths, while still providing sufficient theoretical angular
resolution on large telescopes.

The range of regions with intense star formation stretches from
ultra-luminous galaxies to massive H\,II regions.  Unfortunately, only
the closest regions provide the possibility to identify stars in the
sub-solar mass range.  However, recent studies of starburst galaxies
using the Short-Wavelength Spectrometer (SWS) onboard ESA's Infrared
Space Observatory ISO (Thornley et al. 2000) and complementary 
ground-based near-IR spectroscopy (F\"orster-Schreiber 1998) indicate
that starburst activity occurs on parsec scales in regions of various
sizes and locations which have very similar properties. We conclude
that 30~Doradus and NGC~3603 are representative building blocks of
more distant and luminous starbursts, and use them to address the
following questions:\\
$\bullet$ Is the stellar population produced in starbursts significantly
  different from the stars born in less violent evironments?\\
$\bullet$ Do low- and high-mass stars form simultaneously?\\
$\bullet$ What is the spatial distribution of low- and high-mass stars?\\
$\bullet$ Do low-mass stars form at all?


\section{R\,136}
With $8\times 10^5 M_\odot$ of gas, ionized by $10^{52}$
Lyman-continuum photons/s (Kennicutt 1984), and a surface brightness
of $8\times 10^7 L_\odot$ at its core R136, 30~Doradus is the most
luminous starburst region in the Local Group.  At a distance of
53\,kpc it is possible to spatially resolve most of the stellar
content of its ionizing cluster NGC2070, which contains about 2400 OB
stars (Parker 1993).

\begin{figure}[ht]\centering
\vspace{92mm}
\caption{\label{nic1image} F160W (``H-band'') HST/NIC1 image of 
  R136 in logarithmic scaling.  Pixel scale is $0.\!''043$/pixel for best
  spatial sampling, the FOV is $11''\times 11''\ (2.8 \times
  2.8\mbox{\,pc}^2)$.}
\end{figure}

We observed the 30~Doradus region with NICMOS onboard HST during 26
orbits to establish the H-band luminosity function down to the
faintest practical limit.  Most observations were carried out with the
NIC2 camera which provides a $0.\!''075$/pixel scale ($20''\times
20''$ field of view, FOV).  Although slightly undersampled, we've
chosen this mode because of its superior sensitivity and large field
of view. The observations consist of a $3\times 3\ (56''\times 56'')$
mosaic centered on R136, and two $3\times 1\ (56''\times 20'')$ wings
extending away from the core. All observations were obtained in
MULTIACCUM mode (see Zinnecker \& Moneti (1998) for details) and four
dithered images were obtained for each position to remove bad pixels,
cosmic rays, and the coronographic hole.  Photometry was measured with
DAOphot (Stetson 1987).  No color correction or conversion from F160W
to a ``standard'' H-magnitude has been made; however, comparisons with
ground-based magnitudes for individual stars in R136 (Brandl et al. 1996)
indicate photometric uncertainties of $\le 0.2$~mag (Zinnecker et al.
1998).

The main goal of these observations was to establish the first deep
infrared luminosity function of the 30~Doradus region --- our
``wing~2'' is located Northwest at $r > 15''$ from R136 where crowding
is less severe.  We have detected {\em for the first time} pre-main
sequence stars below $1 M_\odot\ (H\approx 21^{\rm m})$.  The luminosity
function shows an apparent turnover --- probably due to incompleteness
--- at around $H\approx 20^{\rm m}$ (Fig.~\ref{nic2lumi}).  For the range
from $H = 16^{\rm m} \mbox{\ to\ } H=20^{\rm m}$ where our sample is almost
complete we determined the IMF slope $\Gamma$ using the time-dependent
pre-MS mass-luminosity-relation $L_H\sim M^{-1.4}$ for an assumed
cluster age of 2~Myr, based on d'Antona \& Mazzitelli (1994) tracks
and dwarf colors.  We find $\Gamma = 0.84$, which is flatter than the
Salpeter IMF slope of $\Gamma = 1.35$, but very similar to $\Gamma =
0.79$ found in NGC~3603 by Eisenhauer et al. (1998).

\begin{figure}[p]\centering
\includegraphics[scale=0.73]{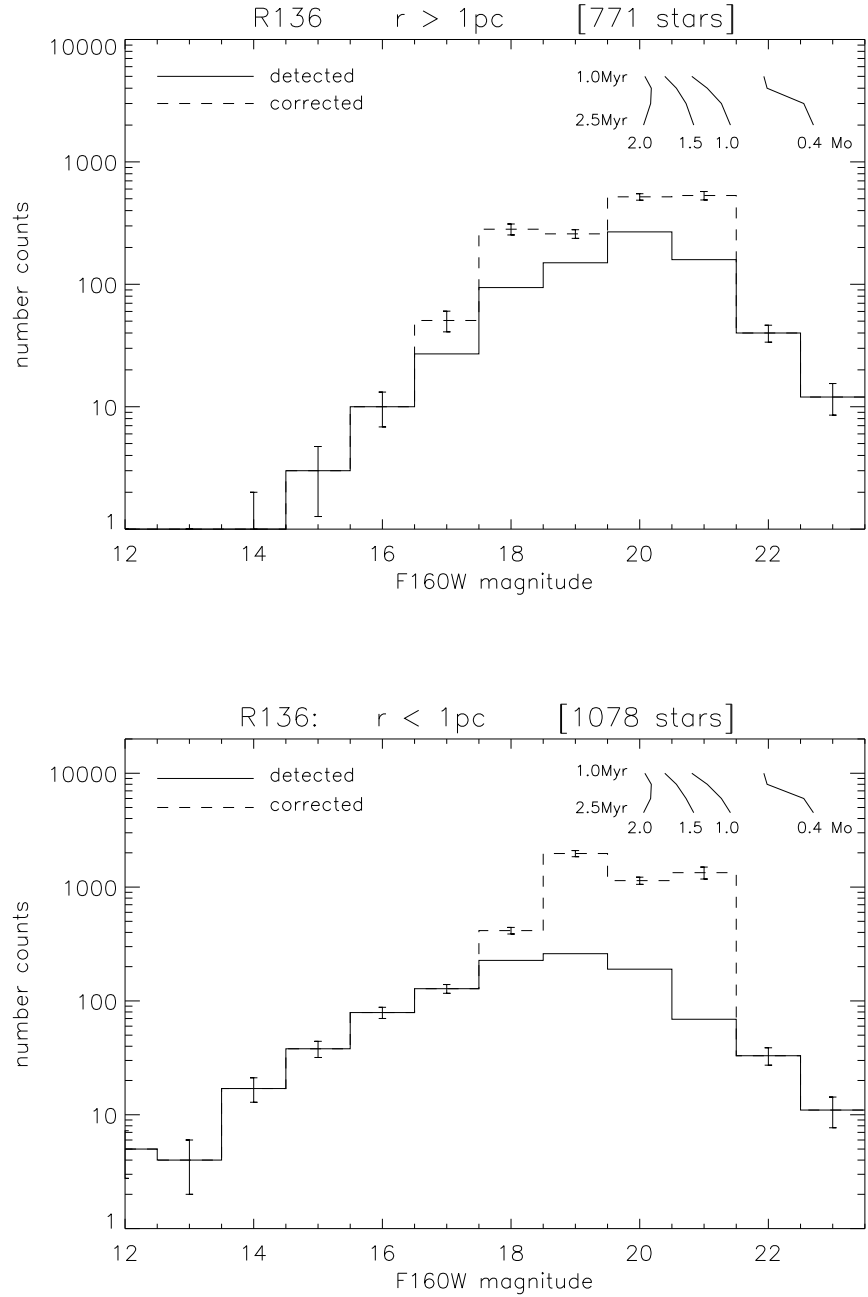}
\caption{\label{r136lumi} F160W luminosity function for about 
  1850 stars with $r > 1$\,pc (top) and $r < 1$\,pc (bottom) within the
  NIC1 $11''\times 11''$ FOV.  The dotted line represents the number
  of actually detected stars whereas the dashed line is corrected for
  incompleteness. The error bars are purely statistical based on the
  number of detected stars.  The upper right part of both diagrams
  indicates the according pre-MS stellar masses of $0.4, 1.0, 1.5, 2.0
  M_\odot$ for ages reaching from 1 to 2.5~Myr (Palla \& Stahler 1999).}
\end{figure}

Most importantly perhaps, our results do not confirm the finding by
Sirianni et al. (2000) mentioned above, i.e., we find no clear
evidence for a significant turnover above $1 M_\odot$.

In addition to the NIC2 data we've also obtained NIC1 ($0.\!''043$/pixel)
images for better spatial sampling on the central $11''\times 11''\ 
(2.8 \times 2.8\mbox{\,pc}^2)$ around R136.  The resulting image is
composed of four exposures of 960s each and shown in
Fig.~\ref{nic1image}.  Due to the significant crowding and extended
diffraction structure in the NICMOS PSF we've chosen a different
analysis approach:
In the first step we analyzed the images with DAOfind (Stetson 1987)
to detect the positions of the brightest sources.  Next we used PLUCY
(Hook et al. 1994) to derive the photometric fluxes for these sources
and produce a map from which the detected sources have been removed.
PLUCY accounts for the full structure of the PSFs within the FOV --
this is important in our case where very faint objects are close to
the diffraction spikes of a bright point source. The PLUCY output map
was used for subsequent source position detections using DAOfind.  The
resulting source list and the original source list were combined and
PLUCY was run again.  These steps were repeated iteratively 4 times.
In total we found about 1850 sources in the field.

Despite this sophisticated approach the non-detection of an existing
source due to crowding is the limiting factor in this analysis.  We
determined the level of incompleteness by randomly adding artificial
stars with a flux distribution following the Salpeter IMF and density
following approximately the cluster profile to the original image and
reanalyzing it through the same procedures as the original image.  To
avoid introducing systematic errors by adding a large number of stars
at once we added only 350 stars at one time and repeated the analysis
two times, which makes it a time consuming procedure.  We determined
the detection probability as a function of apparent magnitude and
distance from the cluster center and applied this correction factor to
the number of stars detected in each luminosity bin.  The result is
shown in Fig.~\ref{r136lumi}.

After correction for incompleteness, the H-band luminosity function at
$r > 1$\,pc ($4''$) around the cluster center but within our
$11''\times 11''$ FOV shows no clear evidence for a truncation down to
about $1 M_\odot$, which is in excellent agreement with our above NIC2
results on the surroundings.  Below $1 M_\odot$ where the uncertainties
from completeness correction become significant no statement can be
made.

Inside the central parsec ($4''$) we detect about a thousand sources.
The luminosity function in the core is flatter compared to the outer
parts of the cluster. For $H\ge 19^{\rm m}$ less than 10\% of the
sources --- whether they exist there or not --- can be detected, and
incompleteness due to crowding does {\em not} permit us to reach below
$2.5 M_\odot$.  The only way to go significantly deeper in the center
is to either wait for NGST and high-order adaptive optics on 8m class
telescopes, or to study targets which are very similar to but closer
than R136.  For now we will follow the latter way.


\begin{figure}[ht]\centering
\vspace{95mm}
\caption{\label{vltbw} Grey-scale version of a $J_{\rm s}$, $H$,
  and $K_{\rm s}$ composite of NGC~3603.  Intensities are scaled in
  logarithmic units; FOV is $3.\!'4 \times 3.\!'4$ ($6.2 \times 6.2$
  parsec$^2$).  North is up, East to the left.  The insert to the
  lower right is a blow up of the central parsec$^2$.  For a color
  version see Brandl et al. (1999). The image also shows the three
  proplyd-like objects that have been recently discovered by Brandner
  et al. (2000); these ``proplyds'' are similar to those seen in Orion
  but about 20-30 times more extended.  About $1'$ south of the
  central cluster, we detect the brightest members of the deeply
  embedded proto-cluster IRS~9.}
\end{figure}

\section{NGC~3603}
The densest concentration of massive stars in the Milky Way --- and
the only massive, Galactic H\,II region whose ionizing central cluster
can be studied at optical wavelengths due to only moderate (mainly
foreground) extinction of $A_{\rm V}\approx 4.5^{\rm m}$ (Eisenhauer
et al.  1998) --- is NGC~3603 (HD97950).  At a distance of only 6\,kpc
(De Pree et al. 1999) NGC~3603 is very similar to R136 (Moffat,
Drissen \& Shara 1994).  With a bolometric luminosity of $10^7
L_\odot$ and more than 50 O and WR stars but only $\frac{1}{50}$ of
the ionized gas mass of 30~Doradus, NGC~3603 is even denser in its
core than R136.  With a Lyman continuum flux of $10^{51}
\mbox{s}^{-1}$ (Kennicutt 1984; Drissen et al. 1995) NGC~3603 has 
about 100 times the ionizing power of the Trapezium cluster in Orion.

\begin{figure}[ht]\centering
\includegraphics[width=5.25in,angle=90,scale=0.52]{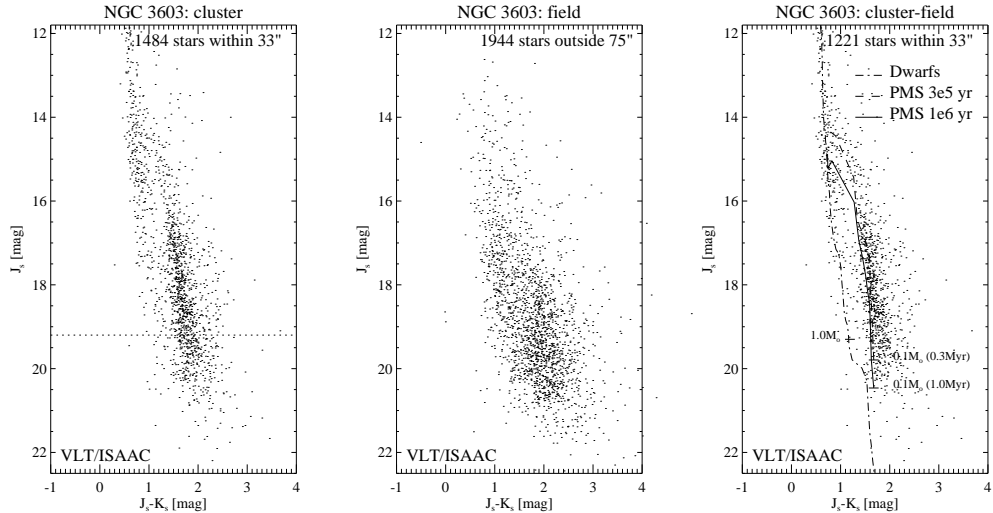}
\caption{\label{vltcmd}{\bf a-c.} 
  $J_{\rm s}$ versus $J_{\rm s} - K_{\rm s}$ color-magnitude diagrams
  of NGC~3603. {\bf a} contains all stars detected in all three
  wavebands within the central $r \le 33''$ (1\,pc) [$1'^2$], {\bf b}
  shows the field stars at $r \ge 75''$ (2.25\,pc) [$6.3'^2$] around
  the cluster, and {\bf c} shows the cluster population within $r \le
  33''$ with the field stars statistically subtracted.  The dashed
  horizontal line in {\bf a} indicates the detection limit of the
  previous most sensitive NIR study by Eisenhauer et al.  (1998).
  {\bf c} also shows the theoretical isochrones of pre-main sequence
  stars of different ages from Palla \& Stahler (1999) and the main
  sequence for dwarfs.  For comparison we've plotted some
  corresponding stellar masses next to the isochrones.}
\end{figure}

We observed NGC~3603 in the $J_{\rm s} = 1.16 - 1.32\mu$m, $H = 1.50 -
1.80\mu$m, and $K_{\rm s} = 2.03 - 2.30\mu$m broadband filters using
the near-IR camera ISAAC mounted on ANTU, ESO's first VLT.  The
service mode observations were made in April 1999 when the optical
seeing was equal or better than $0.\!''4$ on Paranal.  Such seeing was
essential for accurate photometry in the crowded cluster and increased
our sensitivity to the faintest stars.  The details of the data
reduction and calibration are described in Brandl et al. (1999).  The
effective exposure times of the final broadband images in the central
$2.\!'5 \times 2.\!'5$ are 37, 45, and 48 minutes in $J_{\rm s}$, $H$,
and $K_{\rm s}$, respectively.  The images resulting from dithering
are $3.\!'4 \times 3.\!'4$ in size with pixels of $0.\!''074$
(Fig.~\ref{vltbw}).

In order to derive photometric fluxes of the stars we used the IRAF
implementation of DAOphot (Stetson 1987). We first ran DAOfind to
detect the individual sources, leading to $\sim$20,000 peaks in each
waveband.  Many of these may be noise or peaks in the nebular
background and appear only in one waveband.  In order to reject
spurious sources, which may be misinterpreted as a low-mass stellar
population, we required that sources be detected independently in all
three wavebands, and that the maximal deviation of the source centroid
between different wavebands be less than $0.\!''075$.  The resulting
source list contains about 7000 objects in the entire FOV, which were
flux calibrated using faint NIR standard stars from the lists by Hunt
et al. (1998) and Persson et al. (1998) [see Brandl et al. (1999) for
details].

Fig.~\ref{vltcmd}{\bf a} shows the color-magnitude diagram (CMD) for
all stars detected in all 3 wavebands within $r \le 33''$ (1\,pc).
Since NGC~3603 is located in the Galactic Plane and 9 times closer
than R136 we expect a significant contamination from field stars.  To
reduce this contribution we followed a statistical approach by
subtracting the average number of field stars found in the regions
around the cluster at $r \ge 75''$ (Fig.~\ref{vltcmd}{\bf b}) per
magnitude and per color bin (0.5 mag each).  The accuracy of our
statistical subtraction is mainly limited by three factors: first,
low-mass pre-main sequence stars are also present in the outskirts of
the cluster; second, the completeness limit varies across the FOV;
third, local nebulosities may hide background field stars.  However,
none of these potential errors affects our conclusions drawn from the
CMD.

The resulting net CMD for cluster stars within $r \le 33''$ of
NGC~3603 is shown in Fig.~\ref{vltcmd}{\bf c}.  We overlayed the
theoretical isochrones of pre-main sequence stars from Palla \&
Stahler (1999) down to $0.1 M_\odot$. We assumed a distance modulus of
$(m-M)_{\rm o} = 13.9$ based on the distance of 6~kpc (De Pree et al.
1999) and an average foreground extinction of $A_{\rm V} = 4.5^{\rm
  m}$ following the reddening law by Rieke \& Lebofski (1985).

The upper part of the cluster-minus-field CMD clearly shows a main
sequence with a marked knee indicating the transition to pre-main
sequence stars.  The turn-on occurs at $J_{\rm s}\approx 15.5^{\rm m}$
($m\approx 2.9 M_\odot$).  Below the turn-on the main-sequence
basically disappears.  We note that the width of the pre-main sequence
in the right part of Fig.~\ref{vltcmd}{\bf c} does not significantly
broaden toward fainter magnitudes, indicating that our photometry is
not limited by photometric errors.  In fact, the scatter may be real
and due to varying foreground extinction, infrared excess and
different evolutionary stages.  In that case the left rim of the
distribution would be representative of the ``true'' color of the most
evolved stars while the horizontal scatter would be primarily caused
by accretion disks of different inclinations and ages.  Fitting
isochrones to the left rim in the CMD yields an age of only $0.3 -
1.0$~Myr.  Our result is in good agreement with the study by
Eisenhauer et al. (1998) but extends the investigated mass range by
about one order of magnitude toward smaller masses.  

\begin{figure}[ht]\centering
  \includegraphics[width=5.25in]{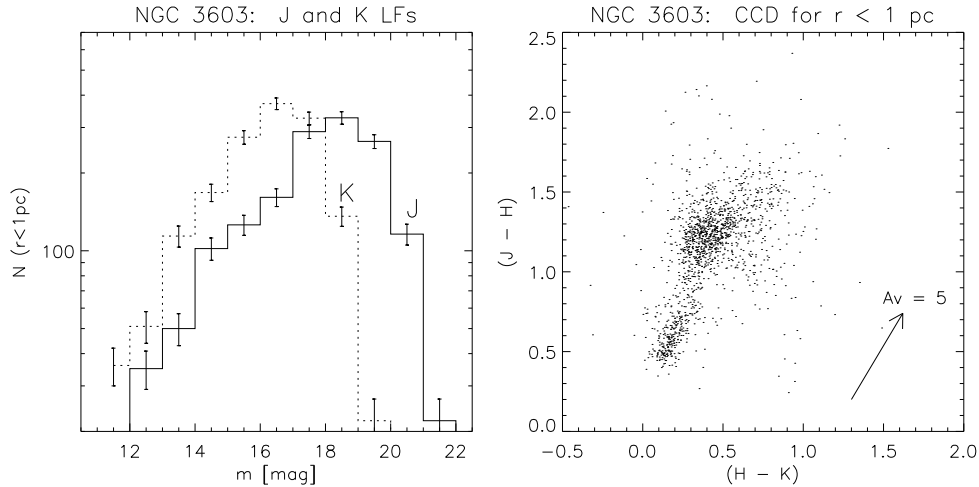}
\caption{\label{newccd} Left: $J_{\rm s}$ (solid) and $K_{\rm s}$-band 
  (dashed) luminosity functions of stars detected at all 3 wavebands
  in the central stellar cluster within $r \le 33''$ (1\,pc). The
  number counts have not been corrected for incompletenes; error bars
  are purely statistical.  Right: Color-color diagram for the same
  1500 stars within 1\,pc of the center.  Redder and younger stars are
  located toward the upper right.}
\end{figure}

Fig.~\ref{newccd} shows the $J_{\rm s}$ and $K_{\rm s}$ luminosity
functions for stars detected in all 3 wavebands with no correction for
incompleteness applied, i.e, toward fainter magnitudes an increasing
number of stars will remain undetected. Thus we cannot say whether the
apparent turnover at $K_{\rm s}\approx 16.5^{\rm m}$ is real or an
observational artifact, but we can state that the mass spectrum is
well populated down to at least $0.1M_\odot$ assuming an age of
0.7~Myr for a pre-MS star with $K_{\rm s}\sim 19^{\rm m}$.

\section{Conclusions}
We have studied the stellar content of the two closest, most massive
starburst regions NGC~3603 and 30~Doradus.  Our studies using
HST/NICMOS and VLT/ISAAC are the most sensitive made to date of dense
starburst cores.  The NIR luminosity function of 30~Doradus shows {\em
  no} evidence for a truncation down to at least $1 M_\odot$, its
stellar core R136 is populated in low-mass stars to at least $2
M_\odot$ and NGC~3603, which is very similar to R136, down to $0.1
M_\odot$.  In a more general picture our findings on starburst cores are:

\newpage
\begin{itemize}
\item{There is no clear evidence for an anomalous IMF.}
\item{The slope of the IMF tends to flattens toward the center.}
\item{There is no evidence for a low-mass cutoff above our completeness 
    limits.}
\item{Sub-solar mass stars do form in large numbers in violent star
    forming regions and are present throughout the cluster.}
\end{itemize}

\acknowledgments{B.B. wants to thank the organizers, in particular
  Thier\-ry Montmerle and Philippe Andr\'{e}, for a very stimulating
  meeting in a beautiful setting.}

\end{document}